\documentclass{article}
\usepackage{graphicx} 
\usepackage{amsmath}
\usepackage{geometry}
\usepackage[]{todonotes}
\usepackage{dsfont}
\usepackage{enumitem}
\usepackage{booktabs}  
\usepackage{subfig}
\usepackage[]{hyperref}
\usepackage{pdfcomment}
\usepackage{tabularx}

\hypersetup{
	colorlinks=true,
	linkcolor=blue,
	citecolor=blue,
	filecolor=magenta,      
	urlcolor=cyan,
	pdftitle={Security Incentivization},
	pdfpagemode=FullScreen,
}

\newcommand{\ie}{i.\,e., }
\newcommand{\R}{\mathds{R}}

\usepackage[nolist]{acronym}
\usepackage{float}    

\title{Security Incentivization: An Empirical Study of how Micropayments Impact Code Security}
\author{Stefan Rass\and Martin Pinzger\and Rainer W. Alexandrowicz\and Georg Sengstbratl\and Johann Glock\and Alexander Lercher\and Fabian Oraze\and Christoph Wedenig
}
\date{\today}

\begin{document}

\maketitle 

\begin{abstract}
Security often receives insufficient developer attention because it does not directly generate visible value, leading to underinvestment in practice. We evaluate a countermeasure by team-level incentives tied to measurable security improvements over time. Our semi-automated mechanism aggregates static analysis findings from Bearer, Detekt, and mobsfscan, computes security issue density, and rewards teams based on the relative improvement ratio across sprints, enabling repeatable, scriptable reporting at scale.

In a controlled course experiment with 84 students across 14 teams, we compared a security-incentivized condition, in which bonus points were linked to security scanner results, against a control condition with an otherwise identical grading scheme. The treatment group achieved significantly lower security issue density overall (beta regression: $\beta = -0.396, p = 0.0342$), indicating improved measurable security under incentivization. After controlling for platform, we observed a marked front-end/back-end disparity, with back-ends showing fewer issues and higher improvement ratios under incentives, highlighting heterogeneous effects across stack layers. Notably, these gains were not the byproduct of inflated code volume, as lines of code increased similarly across groups over time. The measurement pipeline and toolchain proved feasible for scripting and automation, supporting scalable adoption in practice.

Our results suggest that aligning rewards with automated security metrics can measurably improve code security and merit follow-up in professional contexts and longer development lifecycles.
\end{abstract} 

\hypersetup{
    colorlinks=true,
    linkcolor=blue,
    citecolor=blue,
    filecolor=magenta,      
    urlcolor=cyan,
    pdftitle={Security Incentivization},
    pdfpagemode=FullScreen,
    }

    \section{Introduction}\label{sec:intro}
Security has the unfortunate fate of not generating revenue by itself but rather preventing damage at additional cost \cite{Gordon2018EmpiricalEOA,Huang2014OptimalISA,fielder_risk_2018,Anderson2006TheEOA}. 
As such, it does not necessarily ``add'' to the functionality of a system but only protects it from malfunctioning. Consequently, people may take considerably less satisfaction from implementing a security mechanism since the system works before and after, with no visible improvement other than increased robustness and security.

Security must be rooted in the design and implemented concurrently with the functionality throughout the entire development life-cycle.
The problem of reluctance to design a product not only for functionality but also for security has a long history of research, including sophisticated game theoretic analysis using signaling games. Past such work found that companies may not necessarily have natural incentives to invest in security, for example, if customers are unwilling to pay a higher price for a more secure product \cite{RePEc:aea:aecrev:v:85:y:1995:i:5:p:1187-1206}. For an experimental study toward a deeper understanding of why vulnerable code is produced, \cite{fulton_understanding_2022} ran a coding competition to observe the development practices of teams, to find out why and how developers introduce, find, and fix vulnerabilities. It was found that most teams failed to implement correct security mechanisms mainly due to misunderstanding security concepts. Other reasons that effectively hinder developers from writing secure code are known, despite best coding practices such as a secure software development life-cycle (SDL) \cite{ladd_sdl_2011}: These include failure to understand or define implications or security concepts \cite{oliveira_api_2018}, and insufficient resources (time, budget) \cite{blau_behavioral_2017}, among many others \cite{lohrmann_why_2019,nidecki_7_2023,wood_five_2011}. 
AI-powered coding tools, such as Claude Code\footnote{\url{https://www.claude.com/product/claude-code}} or OpenAI Codex\footnote{\url{https://openai.com/codex/}}, hold promise to increase developer productivity, but carry the risk of producing even more insecure code \cite{asare_is_2023, kozak2025aisecurity}.

Consequently, assured quality of security requires not only tools but also developer engagement, motivation, and them taking responsibility for writing secure code. Investments in security may further be due to obligations of independent auditing, (security) standard compliance, or legal obligations (e.g., \cite[\S(64)]{european_comission_horizontal_2024} says that ``Manufacturers should [...] provide security updates to users free of charge'', which is a direct economic obligation in case of inferior security care in product design). Frameworks like the \ac{CC} \cite{ccconsortiumCommonCriteriaInformation2018} or the ISO 27k family \cite{organisationISOIEC270002016} provide well-formalized workflows along which the security of a product, including its production process, can be established. 
These standards depend on developer teams to not only tick checkboxes towards satisfying the minimum requirements that the (security) standard prescribes; for example, if the required security mechanism is implemented according to the standard's requirement, but itself implemented with flaws due to ignorance of security quality beyond what feature is requested. Therefore, practical security will require the willingness to invest the additional effort imposed by a security-by-design paradigm, which might be increased by providing teams with additional payoffs for achieving this goal.

Incentive-based security \cite{huangRepChainReputationBasedSecure2021,liCreditCoinPrivacyPreservingBlockchainBased2018} relies on the hypothesis that people act rationally towards maximizing their personal payoffs. Research has found strong evidence against such a general utility maximization \cite{starmerDevelopmentsNonExpectedUtility2000,tverskyRationalChoiceFraming1989,tverskyFramingDecisionsPsychology1985}, which may be attributed not to a flaw in the general logic of utility maximization, but rather to a mere mistake in how the utilities are modeled. Indeed, research about bounded rationality 
\cite{de_clippel_bounded_2024,starmerDevelopmentsNonExpectedUtility2000} has investigated causes of people not maximizing a presumed utility expressed as a real-valued and continuous function but rather considering multiple dimensions and uncertainty in their decisions. Hence, an easily perceptible incentive to maximize one's utility, for example, an \emph{additional payoff for people who implement security on top of the product's basic functionality}, could help.

Therefore, our research question is: 
\begin{quote}
Does incentivizing security improvements through bonus (e.g., grade points) lead to lower security issue density than equivalent incentives for general code quality?
\end{quote}

To answer this question, we design a mechanism to measure security contributions, together with an experiment to verify it. The mechanism employs an automated tool chain to quantify security improvements over time by changes in the security issue density. From these changes, i.e., improvements, we define rewards to incentivize developer teams. The incentive mechanism was evaluated in an empirical study with students of the software engineering class at the University of Klagenfurt, demonstrating that such additional incentives can increase security. 

Our contribution comprises: 
\begin{itemize}
	\item A semi-automated mechanism to measure security contributions in teams, which we empirically pre-tested.
	\item An experiment with students of the software engineering class at the University of Klagenfurt to empirically support the efficacy of the incentivization for better security.
\end{itemize}

In Section~\ref{sec:related-work}, we provide an overview of previous approaches to use incentivization for increasing software security. Section~\ref{sec:incentive-counting-mechanism} discusses mechanisms by which incentivization may be engendered and describes a pre-study we performed to gain insight of the feasibility of such an approach. Section~\ref{sec:experiment} describes the details of the present experiment and Section~\ref{sec:results} shows the outcomes and their discussion.      \section{Related Work}\label{sec:related-work}

Gamification is one of the most commonly applied technique to increase motivation in various contexts, like work (e.g., ``employee of the month'', ``best-performing teams'', etc.) \cite{reiners_studying_2015}. Using game-based rewards is also applied to increase survey participation \cite{dejonckheere_real-time_2024}, although with some caveats. The underlying idea is leveraging peoples motivation to ``outperform'' others. Such \emph{extrinsic motivation} means doing an activity for a separable outcome, like rewards, recognition, status, avoiding punishment, or social comparison. It is complementary to \emph{intrinsic motivation}, i.e., doing an activity because it is inherently interesting or enjoyable for curiosity, mastery, or challenge for its own sake. The latter is often found behind people's engagement in open source software projects. Since intrinsic motivation strongly depends on character traits, it is harder to trigger externally than extrinsic motivation. Possibly problematic for our purposes are social comparison-based techniques (e.g., gamification), since people can reach the relative top position either by increasing their efforts (positive and desired), or by downplaying other's performance (negative, undesired up to unethical). Our focus in this work is hence on extrinsic motivation by individual performance, explicitly avoiding any relative or comparison-based judgement. This is extrinsic-external motivation. Further variants like extrinsic-introjected motivations (doing things for pride or to avoid shame), or extrinsic-identified motivations (doing something because it aligns with personal values or goals) for security will naturally serve the purpose, and hence need no extra intervention.

The challenge of incentivizing software developers to care more for security has long been recognized. The early work of \cite[p.43]{augustDesigningUserIncentives2014} states that ``\textit{the design of an incentive-based approach to improve cybersecurity is a difficult task because the level of risk that realizes on a given system or network is a complex outcome of the behaviors of many stakeholders: government, critical infrastructure providers, technology producers, malicious (`black hat') hackers, and users}''. This motivates a change to a more intrinsic mechanism that quantifies security contributions exclusively on the actions of the developers, which we can measure (unlike all other ``variables'' mentioned above).

The idea to declare security as a goal in its own right is contrasted by the dual approach of transferring risk. In special contexts such as IoT, this can mean mechanisms of transferring risk to third parties (such as insurance) \cite{adatEconomicIncentiveBased2018}. Early game-theoretic treatments of company liabilities for insufficient security (modeled by signaling games) have shown that investments into security can correlate with pricing in a way that can create even the opposite incentive of \emph{not} investing in security \cite{RePEc:aea:aecrev:v:85:y:1995:i:5:p:1187-1206}. The work of \cite{haldermanStrengthenSecurityChange2010} provides eloquent discussions about the need for the right incentives and argues for incentivization using transparency and liability mechanisms. In light of the aforementioned research, especially liabilities may game-theoretically induce unwanted effects, so the problem seems to require other mechanisms (one of which we propose in this work). Possible negative effects of market-based incentives for security were also found by \cite{eetenEconomicsMalwareSecurity2008}. 

Incentivization mechanisms for security do not need to rely on the developers themselves, but can equally well root in the user group \cite{palombo_ethnographic_2020}. Bug bounty programs are a common instance, offering rewards for security flaw reporting. Our work looks into how a likewise incentive for developers can a priori mitigate flaws during the development already, instead of offering rewards for finding security flaws in a deployed product (bug bounty). The work of \cite{augustNetworkSoftwareSecurity2006} provides such a mechanism, discussing different patching strategies, and \cite{garayRationalProtocolDesign2013,gordonRationalSecretSharing2006,kawachiGeneralConstructionsRational2017,huangRepChainReputationBasedSecure2021,liCreditCoinPrivacyPreservingBlockchainBased2018} use incentivization mechanisms to encourage honest behavior in cryptographic protocols and blockchains. Similarly, \cite{liuIncentivebasedModelingInference2005a} uses non-cooperative game theory to study incentives for attackers. The most recent work of \cite{sunWhoRealHero2023a} pushes this further by systematically identifying strong contributors in a collaborative software project; a mechanism with a direct application to our incentivization scheme.
	
A finegrained mechanism proposed in \cite{rass_incentive-based_2023} and based on the Shapley value \cite{shapley_valu_1951, roth_shapley_1988}, can break down the team accomplishments into contributions by individual team members and reward them accordingly. We review this mechanism's principles in Section \ref{sec:reward-assignment} briefly, but will -- to avoid creating undesired intra-team-dynamics that could lead to biases in the study -- reward all team members equally. Further applications of the Shapley value have been described in \cite{rass_incentive-based_2023} and the software alliances \cite{lvResearchProfitDistribution2013} and to share threat intelligence \cite{xieImprovedShapleyValue2020}. 

Our work adopts a simpler reward mechanism that targets a team, rather than an individual, since the attribution of a contribution to a specific person induces considerably more uncertainty than linking the contribution to a team. We explain the reasons in Section \ref{sec:pre-study}. 
\section{Incentivizing Security Contributions}\label{sec:incentive-counting-mechanism}

Security is well supported in development environments by numerous tool integrations, but in any case, leaves the final effort up to the developer to manually use the tools and enhance the code towards more security. While this imposes a considerable effort of time and resources, the developer's perceived benefit for the development goals, e.g., requested features, bugfixes, or similar, may fall behind when it comes to security. Even worse, if additional security mechanisms negatively impact usability, the incentive may even be reverted into \emph{not} investing into security for the sake of simplicity. This happened, for instance, for the early versions of the \ac{ROS}, that were initially built without any security, which eases integration and using ROS, but induced the inevitable, and demonstrated, risk of malfunctions and misuse \cite{dieber_security_2017}.

The narrative of \emph{incentive-based security} is to make it a developer's own interest to not only properly implement the desired functionality. It additionally aims to make systems secure beyond what compilers, e.g., like for Rust, or a modern development environment like IntelliJ, Visual Studio, etc., with their built-in security analysis features may do. Note, while recent approaches showed to have improved detecting \cite{kaniewski2025review} and even remediating security issues \cite{repairbench}, it is still the task of developers to filter false positives, and perform and validate security fixes.

To this end, we propose a mechanism to measure and feed back to the developer the added value of code changes that increase security, thereby incentivizing him/her.

\subsection{Security Incentivization Mechanism}\label{sec:the-mechanism}
We propose to measure the code's quality in terms of security, and reward a developer for changes of the quality towards better security. Figure \ref{fig:procedure} summarizes the procedure. 

\begin{figure*}
	\begin{tabularx}{0.95\textwidth}{c|l|X|p{2.5cm}}
		\multicolumn{2}{c|}{Procedural Step} & Description & Details\tabularnewline
		\hline 
		1 & Scan & Analyze a software system for \textbf{security issues} and \textbf{count}
		them. & Section \ref{sec:counting}\tabularnewline
		\hline 
		2 & Estimate change & Determine a \textbf{trend towards more or fewer issues} over time. This can
		mean a simple absolute or relative difference between past and current
		issue counts, up to sophisticated time series modeling and forecasting,
		though the latter is beyond the scope of this work.  & we use Equation \eqref{eq:issue-density} for quantifying changes of security\tabularnewline
		\hline 
		3 & Pay rewards & Assign a bonus to the team or the developer if the issues become fewer
		(on average) over time. This is the \textbf{core part} of the mechanism, as
		this aspect creates the developer's incentive to invest time and efforts
		into security, while \emph{preserving} (at best) the existing functionality
		only. & generally: Section \ref{sec:reward-assignment}, specific for our
		experiment: Section \ref{sec:the-bonus-system}\tabularnewline
		\hline 
		4 & Repeat & Go back and \textbf{repeat from Step 1}, in intervals that are meaningful for
		the project, e.g., sprints. & Section \ref{sec:experiment}\tabularnewline
		\hline 
	\end{tabularx}
	\caption{Security Incentivization Procedure}\label{fig:procedure}
\end{figure*}

\vspace{0.5\baselineskip}
To test these generic steps for \emph{practicality} (feasibility in practice) and \emph{fitness for use} (accomplishing what is needed), we conducted a preliminary study in Section \ref{sec:pre-study}. Subsequently, we evaluated the \emph{fitness for purpose}, i.e., efficacy of the incentivization mechanism itself, in a second experiment described in Section \ref{sec:experiment}, and discuss its efficacy in Section \ref{sec:reward}.

\subsection{Counting Security Issues}\label{sec:counting}
Towards a practical implementation of a security quality scoring in the above mechanism, we prefer automated tools to identify and count security issues. Their availability depends on the programming language. 

From the scanners that classify as feasible for the application (our concrete choices are given in Section \ref{sec:pre-study-design}), we sum up the issue counts at time $t$ in the value $I_t$. Duplicate findings of two scanners may be retained, based on the intuition that an issue that several scanners find may be either more severe or easier to fix, which in both cases, merits its prioritization for fixing. Double (or multiply) counting the issue, in turn, puts more weight on the fact that this issue is fixed later. While much research is devoted to leveraging \acp{LLM} to process unstructured information, especially from \ac{SAST} tools \cite{lykousas_potential_2024}, 
the use of such methods entails a need to verify the security alert before fixing it, anticipating hallucinations or incorrect replies from the \ac{LLM}, and thus can increase the burden on the developer undesirably.

To measure the change of security quality, we consider \emph{security issue density}, defined as the number of issues per \ac{LOC} at the current release time $t$, $Q_t = \frac{I_t}{\text{LOC}_t}$. Taking the density instead of the absolute issue counts rates the current security regardless of the absolute code size and prevents developers from being penalized for more complicated code or many LOC, thus avoiding unfairness in this respect. To then capture the improvements or deteriorations of security, we compute the \emph{relative change} in security issue density 
\begin{equation}\label{eq:issue-density}
	\Delta Q = \frac{Q_{t_{i+1}}}{Q_{t_i}}\in[0,\infty)
\end{equation}
between release times $t_i$ and $t_{i+1}$ or sprints.

From $\Delta Q$, we define a reward awarded to developers as a monotonously decreasing function $R:[0,\infty)\to\R^+$, so that lower values of $\Delta Q$ leads to higher rewards; materializing the logic that a lower issue density at a later release, relative to higher issue density before, indicates a security improvement. Our mechanism rewards improvements whenever $\Delta Q<1$, but does not punish a developer if $\Delta Q\geq 1$.

   	\subsection{Linking Contributions to Teams or Individuals?}\label{sec:reward-assignment}
    
    The reward scheme in \cite{rass_incentive-based_2023} was proposed from first principles (P1\ldots P4) to pay bonuses to developers individually based on the following intuition: let $R(\Delta Q)$ be a monetary or immaterial reward (the latter was used in our experiment in Section \ref{sec:experiment}), which is awarded to individual developers according to the following rules:
\begin{enumerate}[label=P\arabic*]
	\item \label{lbl:null-player} A person who has not provided any security contributions does not receive a reward.
	\item \label{lbl:linearity} A person whose security contributions are beneficial in several projects (e.g., fixing a library shared by several applications) receives the total of rewards from all applications where the security fix applies.
	\item \label{lbl:fairness}  The reward is independent of the person's role in the team (fairness).
	\item \label{lbl:full-payment} The team reward $R(\Delta Q)$ is fully shared among all developers that contributed to security.
\end{enumerate}

These principles directly correspond to the defining axioms of the Shapley-value \cite{roth_shapley_1988}, but the so-designed mechanism strongly depends on how software developers commit code changes to the repository.
In particular, developers might commit incomplete code changes that may not even compile or run, likely raising more issues if the scan is triggered upon every commit. Furthermore, merging of pull requests might squash commits into a single commit, thereby changing the authorship of commits. Finally, even if the original author information is retained, the results of a security scan are not easy to link accurately to individual developers' code contributions. Multiple developers might contribute to fixing a security issue while only one developer commits the code changes.

Therefore, in this work, the security rating, i.e., computation of $\Delta Q$ and assignment of rewards $R(\Delta Q)$ is made under the following conditions only:
\begin{itemize}
	\item The code change was merged into the \texttt{main} branch of the repository. We assume that other branches are dedicated to the development of new features or other versions of the same system.
	\item The reward $R(\Delta Q)$ is paid equally to all team members, to avoid internal competition among team members, and to let the whole team benefit from the member's individual talents (coding, security, etc.) for the good of the whole team. While this is against principle \ref{lbl:null-player}, we argue that different talents, when combined in complementary ways, merit a mutual benefit even without a direct contribution to the goal (whether it is adding a feature or adding to security).
\end{itemize}
Obviously, this mechanism satisfies the other properties \ref{lbl:fairness} and \ref{lbl:full-payment}, while property \ref{lbl:linearity} is easy to implement in addition, if rewards for code changes are measured across several branches or projects.  
\subsection{Pre-Study of the Issue Counting Mechanism}\label{sec:pre-study}

During the winter term of 2024, spanning from October to January 2025, we conducted a preliminary study in a course on software development at the Johannes Kepler University Linz, to explore the concept of scanning code repositories for identifying issues and measuring security contributions. The goal was to proactively uncover potential problems and challenges with the mechanism to determine the number of issues automatically.

\subsubsection{Pre-Study Design}\label{sec:pre-study-design}
Before running the pre-study with the student teams, we tested multiple scanners to ensure a smooth and reproducible setup for the pre-study course. 

We currently considered Kotlin and Java as two major languages on mobile platforms (e.g., Android). For these, we identified three suitable \ac{SAST} tools for our purpose, which are \texttt{Bearer}\footnote{\url{https://github.com/Bearer/bearer}, retrieved: Jan.16th, 2026}, \texttt{Detekt}\footnote{\url{https://github.com/detekt/detekt}, retrieved: Jan.16th, 2026}, and \texttt{mobsfscan}\footnote{\url{https://github.com/MobSF/mobsfscan}, retrieved: Jan.16th, 2026}. Common to these three is their usability for scripting, and reliable automated parsing of the resulting reports, to extract the individual issue counts. 

In addition to these three, we also evaluated Snyk \cite{snyk_limited_snyk_nodate}, Aikido \cite{aikido_security_bv_aikido_nodate}, HCL AppScan \cite{hclsoftware_ai-powered_2025}, Semgrep \cite{semgrep_inc_semgrep_2025}, AppSweep \cite{guardsquare_appsweep_nodate}, and SonarQube \cite{sonarsource_sarl_sonarsourcesonarqube_2025}. The pre-test was performed on open-source projects as well as selected course projects from last years courses. This enabled us to verify stability, rule coverage and most importantly how well the tools can be integrated in a CI process, such that students have an easy access to the results.
As part of the study, we provided student groups tasked with software development a ``free service'' to perform security scans on their code. The results of these scans were shared with the participants, highlighting any issues detected. In exchange, we gained access to their code repositories (Git repositories), enabling us to test various implementations of scripting-based code scanners.
These scanners were used to auto-generate reports and quantify the number of identified issues. The issues were reported back to the students, but without any influence on their course grading (for ethical reasons laid out in Section \ref{sec:pre-study-ethics}). To validate the scanners, participants were asked to fix specific vulnerabilities or intentionally leave them unresolved. This process ensured the scanners could accurately detect vulnerabilities both before and after changes. The study was designed to benefit both parties: participants received regular security checks to improve their code, while researchers gathered data to optimize the measurement mechanism for the upcoming experiment.

\begin{figure*}
	\centering
	\includegraphics[width=\textwidth]{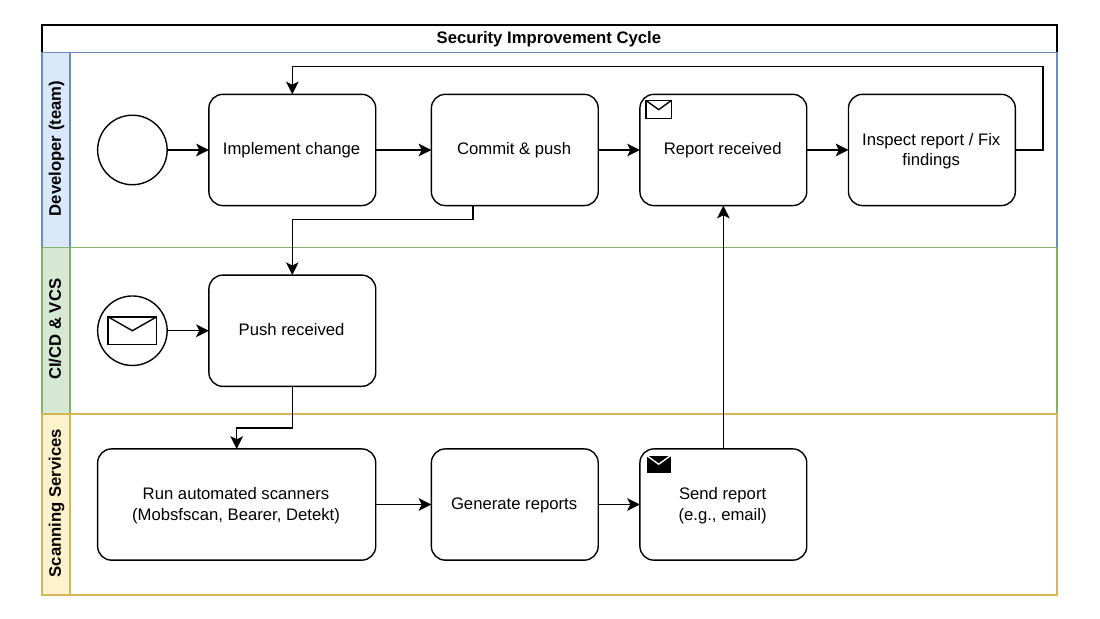}
	\caption{Security improvement cycle with automated scanning and developer feedback. After a push to the repository, the CI/CD pipeline is triggered and scanning services are executed. The resulting reports are collected and sent to developers, who inspect findings and apply fixes. Committing the changes triggers the pipeline again, continuing the feedback loop. This mechanism is detailed in Section~\ref{sec:the-mechanism} and forms the basis of the preliminary feasibility study (Section~\ref{sec:pre-study}) and the efficacy experiment (Section~\ref{sec:experiment}).}\label{fig:issue-counting-mechanism}
\end{figure*}

For further verification, additional manual checks were performed on the scanner outputs. While most findings were directly related to security, some were not. Nevertheless, issues related to code quality were included in the overall issue count $I$. This design decision was based on the rationale that poor code quality could potentially evolve into significant security vulnerabilities over time, either directly or by complicating code maintenance and correction. 
Figure \ref{fig:issue-counting-mechanism} illustrates the mechanism of this preliminary study, where the feedback after the manual check was sent back to students over email for their (further unconstrained) use. This provided valuable insights into the feasibility and challenges of using automated tools for security measurement, with the findings for the setup listed in Section \ref{sec:pre-experiment}.

\subsubsection{Implementation and Results of the Pre-Study}\label{sec:pre-experiment}

To ensure comparability, we enforced \texttt{git} practices whereby development teams regularly pushed their changes to the \texttt{main} branch. We treated main as a proxy for the current production state; accordingly, it was expected to compile cleanly and contain fewer security vulnerabilities than transient development branches.
Alternative tools (tested in Section \ref{sec:pre-study-design}) either identified too few weaknesses in our codebases or were cumbersome to integrate for the study setup, so we ultimately limited the pre-study tooling to \emph{Bearer}, \emph{Detekt}, and \emph{mobsfscan}. In addition to the quality of the findings, it was important to us that the tool was easy for the students to use. One of the major issues with SAST tools is that they all have different output formats or dashboards. To resolve this issue a new industry standard the Static Analysis Results Interchange Format (SARIF~\cite{fanning_static_2025}) emerged. The SARIF format is a standardized JSON schema that is easy to  view with VS-Code extensions or process programmatically. Our selection of scanners is using the SARIF format. The pre-study setup consisted of a Jenkins build server with pipelines that were executed after each commit on the main branch. Jenkins would then create a report for each scanner, send it to the students via email, and log the reports in a database. For stability testing, we also set up a Jenkins pipeline for a popular open-source Android app. Table \ref{tab:pre-study} shows the team sizes and programming language in the pre-study, on which the scanners underwent a deeper evaluation for feasibility in an automated tool chain for security issue counting. 

\begin{table}
	\centering
	\caption{Teams Pre-Study}
	\label{tab:pre-study}	
    \begin{tabular}{cccccc}\toprule
		ID & Team Size & Language & Bearer & Detekt & mobsfscan\\\midrule
		1 & 2 & Kotlin & 0 & 69 & 14 \\
		2 & 1 & Swift & 0 & 0 & 8\\
		3 & 3 & Java & 0 & 3 & 8\\
		4 & 1 & Kotlin & 0 & 67 & 18\\\bottomrule
	\end{tabular}
\end{table}

Since the LOC is a important metric for us in order to calculate the issue density we have to understand how our selected scanners process the source code. Detekt is a code quality scanner for the Kotlin programming language, the open source version of Bearer can perform SAST for the Java programming language. mobsfscan is a static analysis tool that is capable to scan Java and Kotlin code with a special focus on mobile applications, e.g. Android and iOS, which have special rules that includes also the \texttt{AndroidManifest.xml} file. The calculation of the LOC was therefore based on the source files our selected scanners can process therefore we counted the lines without comments and block comments for Java, Kotlin source files and in addition to that the \texttt{AndroidManifest.xml} files.

The pre-study provided no evidence against the practicality of an automated security issue counting using equation \eqref{eq:issue-density}, so we proceeded with the main experiment. 

\section{Experiment Design}\label{sec:experiment}

    Building on the pre-study (Section~\ref{sec:pre-study}), we designed a controlled experiment (leveraging the feedback and improvement cycle from Figure \ref{fig:issue-counting-mechanism}) to answer the following research question from Section \ref{sec:intro}.

The study involved 84 Bachelor's students enrolled in the software engineering lab course at the University of Klagenfurt during the summer term of 2025. The students were organized into 14 development teams of 4-7 students to complete a semester-long programming project. We assigned the teams to one of two conditions: a control group (CON, 6 teams, 32 students), incentivized for general code quality, and a treatment group (SEC, 8 teams, 52 students), incentivized for reducing security issues. All participating students were fourth-semester undergraduates who had completed introductory courses
in programming and algorithms but had not received formal security education. Some students had concurrent or prior industry experience, but the majority did not.

The remainder of this section describes the course and project setup (Section~\ref{sec:study-setting}), participants and group formation (Section~\ref{sec:participants}), incentive conditions (Section~\ref{sec:conditions}), procedure and timeline (Section~\ref{sec:procedure}), and experiment controls (Section~\ref{sec:controls}).

\subsection{Course and Project Setup}\label{sec:study-setting}

Each team developed a networked multiplayer adaptation of an existing board or card game for the Android platform, selecting their game subject to instructor approval. The only strict architectural requirement was to implement client-server communication via a dedicated server. Most teams additionally implemented session management (e.g., game lobbies). We retained the game domain from previous course iterations. For the purposes of this study, the domain offers a comparable attack surface across projects while avoiding highly sensitive data (e.g., medical or financial information) that would introduce variability in security posture.

Each team maintained two separate codebases in independent Git repositories: an Android client application (the front-end, implemented in Kotlin) and a server application (the back-end, implemented in Java or Kotlin at the team's discretion). There were no requirements regarding using a database or persistence layer, though some teams chose to incorporate them.

\subsection{Participants and Group Formation}\label{sec:participants}

Students self-selected their course section based on scheduling preferences. Enrollment occurred before we announced any bonus system. Instructors had been assigned to sections before enrollment opened, so section sizes could not be anticipated. After enrollment closed, we assigned sections to experiment conditions (CON vs. SEC) subject to two constraints: (a)~Instructor~2 should teach in both conditions and (b)~group sizes should be as balanced as feasible given the natural enrollment distribution. Table~\ref{tbl:group-sizes} shows the resulting assignment.

At the start of the semester, students completed a questionnaire covering their study program, overall programming experience, Android-specific experience, and project experience.
We informed the students that the questionnaire would be relevant for the team assignment, but did not disclose specific assignment procedures.
Instructors used the self-reported experience data to compose teams with a balanced mix of skill levels.
This yielded 6 teams in CON (32 students, average team size of 5.3) and 8 teams in SEC (52 students, average team size of 6.5).

\begin{table}[htbp]
    \centering
	\caption{Distribution of participants across sections and experimental conditions}\label{tbl:group-sizes}
	\begin{tabular}{|c|c|c|c|c|}
		\hline 
		Section & Instructor(s) & Condition & Students & Teams\tabularnewline
		\hline 
		\hline 
		I & 1 & CON & 20 & 4 \tabularnewline 
		\hline 
		II & 2 & CON & 12 & 2 \tabularnewline 
		\hline
		\multicolumn{3}{|c|}{\textbf{CON Total}} & \textbf{32} & \textbf{6} \tabularnewline
		\hline
		\hline
        III & 2 & SEC & 33 & 5 \tabularnewline 
		\hline
        IV & 3, 4 & SEC & 19 & 3 \tabularnewline 
		\hline
		\multicolumn{3}{|c|}{\textbf{SEC Total}} & \textbf{52} & \textbf{8} \tabularnewline
		\hline
	\end{tabular}
\end{table}

\subsection{Incentive Conditions}\label{sec:conditions}\label{sec:the-bonus-system}

Each student could earn up to 200 individually assessed base points and up to 20 bonus points. The two experimental conditions shared identical base grading, passing requirements, and maximum bonus. They differed in a single respect: the metric that determined the bonus points. Figure~\ref{fig:synopses} summarizes the course synopsis as presented to all students.

\begin{figure*}[htbp]
	\small
	\fbox{
		\begin{minipage}{\dimexpr\textwidth-2\fboxsep-2\fboxrule\relax}
			\begin{itemize}
				\item Implement the assigned project in a group of 5--7 members across 3 agile (Scrum-like) sprints of roughly 1--1.5 months each.
				\item Grading criteria:
				\begin{enumerate}
                    \item Functionality and software quality (using SonarQube quality gates).
                    \item Application of software engineering techniques (testing, CI).
                    \item Use of agile development practices (Scrum events, backlog management).
				\end{enumerate}
				\item Scoring: 0 to 200 points (10 + 50 / 70 / 70 points per sprint).
                \begin{itemize}
                    \item 10 points before sprint 1 for an individual admission assignment.
                    \item 15 / 20 / 20 points in sprint 1 / 2 / 3 for grading criterion 1.
                    \item 15 / 20 / 20 points in sprint 1 / 2 / 3 for grading criterion 2.
                    \item 20 / 30 / 30 points in sprint 1 / 2 / 3 for grading criterion 3.
                \end{itemize}
				\item Individual contributions must be committed to the team's Git repositories under each participant's name. Only contributions that are merged into the main branch at the end of each sprint are considered for grading.
                \item Bonus: Up to 0 / 10 / 10 additional points can be awarded in sprint 1 / 2 / 3 for improving \textbf{security scanner results} beyond the minimum requirement to pass the course.
			\end{itemize}
	\end{minipage}}
\normalsize
	\caption{Course synopsis provided to the security-incentivized group (SEC). The control group (CON) received identical instructions, with the sole exception that the bonus criterion read ``Sonar scanner results'' instead of ``security scanner results'' (highlighted in bold).}\label{fig:synopses}
\end{figure*}

Both conditions faced identical passing requirements, unchanged from past cohorts: (i)~a minimum of 120 individual points, (ii)~at least one completed feature per student (\ie a vertical slice covering user interface, game logic, and networking), and (iii)~team-level SonarQube quality gates (maximum 100 issues total across both repositories, at least 70\% test coverage for game logic). Notably, neither group faced a baseline security requirement.

We distributed the 200 base points across an individual admission assignment (10 points, due before the Sprint 1) and three agile sprints (50, 70, and 70 points respectively). Within each sprint, we assessed base points \emph{individually} based on each student's contributions across three criteria: (1)~functionality and software quality, (2)~application of software engineering techniques (e.g., testing and continuous integration), and (3)~use of agile development practices (e.g., Scrum events and backlog management).

In addition to the base points, each student could earn up to 20 bonus points --- enough to improve the final mark by one full grade. The bonus was available only in Sprint~2 and Sprint~3 (up to 10 points each). Sprint~1 offered no bonus. We determined the bonus at the team level: regardless of which members contributed to the improvement, every member received the same amount. Section~\ref{sec:reward-assignment} provides further motivation for this team-level reward structure.

The sole difference between conditions was the metric driving the bonus. In the control condition (CON), students earned bonus points for improving SonarQube code quality metrics beyond the passing threshold. We calculated the bonus according to Equation~\ref{eq:quality-reward}:
\begin{equation}\label{eq:quality-reward}
    R_Q = \mathtt{round}\left(10 \cdot \frac{100 - I_{\text{total}}}{100}\right)
\end{equation}
where $I_{\text{total}}$ is the sum of remaining SonarQube issues across the Android client (maximum 50 issues to pass) and the server component (maximum 50 issues to pass). This bonus system had been used in previous course iterations, and we communicated the exact formula to CON students.

In the treatment condition (SEC), students earned bonus points for reducing security issue density as measured by Bearer, Detekt, and mobsfscan (the tools described in Section~\ref{sec:counting}). We calculated the bonus according to Equation~\ref{eq:reward}:
\begin{equation}\label{eq:reward}
    R(\Delta Q) = \mathtt{round}(\min[10, 10 \cdot (1 - \min(\Delta Q, 1))])
\end{equation}
where $\Delta Q = Q_{t_{i+1}} / Q_{t_i}$ is the ratio of security issue density between consecutive assessment points (defined in Section~\ref{sec:procedure}). We did not communicate the exact formula to SEC students, as it had not been finalized at the time of the course briefing. Instead, we told them that bonus points would be awarded for improving their security scanner results and that eliminating all flagged security issues would yield the full bonus.

\subsection{Procedure and Timeline}\label{sec:procedure}

We divided the semester into three agile sprints, assessing each team's code at sprint boundaries ($t_1$, $t_2$, $t_3$). Sprint~1 (weeks 12--15 of 2025) covered application architecture and networking infrastructure. Sprint~2 (weeks 15--20) added core game mechanics, producing a playable game. Sprint~3 (weeks 20--26) focused on bug fixing, UI polish, and optional features, culminating in a joint presentation across all four sections in the last week of the lab course (week 26). Exact assessment dates varied by section based on scheduled meeting days (Monday, Tuesday, or Thursday), and instructors occasionally granted brief extensions.

At each assessment point, we scanned the source code in the \emph{main} branch of all repositories using Bearer 1.49.0, Detekt 1.23.8, and mobsfscan 0.4.5, following the scan--audit--manual-check procedure described in Section~\ref{sec:counting} (Figure~\ref{fig:issue-counting-mechanism}) and validated in the pre-study (Section~\ref{sec:pre-study}). We counted lines of code for each repository to compute security issue density. The scanning tools are open-source, so students could also run them independently, though the extent to which they did so was not recorded. In the course session following each assessment point, instructors discussed team progress, SonarQube reports, and security scan reports with each team to ensure the bonus mechanism remained salient.

Lines of code were computed with a simple Python script that 
iterated over \texttt{.java} and \texttt{.kt} source files and excluded both line (\texttt{//}) and 
block (\texttt{/* ... */}) comments.

\subsection{Experiment Controls}\label{sec:controls}

Our design controlled for two primary classes of potential confounds: unequal treatment between conditions and bias in group assignment and evaluation.

To isolate the effect of the incentive target, we ensured that the two conditions differed only in the metric that determined the bonus. Both offered the same maximum bonus (20 points over two sprints). The two reward formulas differ structurally, reflecting the deliberate retention of the established CON formula alongside a newly developed SEC formula (see Section~\ref{sec:conditions} for details). Because the SEC formula computes improvement as a ratio between consecutive assessments, it requires a prior measurement point. Neither condition could therefore earn bonus points in Sprint~1, which provided the initial measurement~($t_1$). Both conditions received security scan reports and SonarQube results at every assessment point, with one set of results tied to the bonus and the other presented as informational. Observed differences in outcomes therefore reflect the incentive structure, not differential access to information.

To prevent bias in group assignment and evaluation, we separated enrollment from condition assignment, maintained blinding, and kept evaluation independent of instruction. Because students enrolled before we announced the bonus system and we assigned conditions only after enrollment closed (Section~\ref{sec:participants}), students could not self-select into experimental conditions. We did not inform students that the course involved a comparative study, and the written briefing materials differed by a single phrase (``Sonar scanner results'' vs.\ ``security scanner results,'' see Figure~\ref{fig:synopses}). We discuss the ethical aspects of non-disclosure in Section~\ref{sec:experiment-ethics}. Instructor~2 taught sections in both conditions, and all sections received identical written instructions (Figure~\ref{fig:synopses}) with automated outcome measurement. We discuss residual instructor confounding in Section~\ref{sec:threats}. Researchers, who were not involved in teaching, performed the security scans and statistical analysis, reducing the risk of experimenter bias.

\section{Results}\label{sec:results}

    \renewcommand{\arraystretch}{1.3}

Before turning to the details, we consider the potential impact of the two programming languages, Java and Kotlin, used in the projects. Table~\ref{tab:lang} shows the almost perfect overlap of back-ends implemented in Java and front-ends in Kotlin. Only three teams used Kotlin for implementing the back-end; no team used Java for implementing the front-end. Therefore, the subsequent analyses refer only to the distinction of back-end vs. front-end layers. The results also apply, by and large, to the distinction between the two programming languages.

\begin{table}
\centering
\caption{Number of teams using Java or Kotlin to implement the back-ends and front-ends.}
\label{tab:lang}
\begin{tabular}{rrrr}\toprule
   & Java & Kotlin & Sum \\ \midrule
  back-end & 11 & 3 & 14 \\ 
  front-end & 0 & 14 & 14 \\ \midrule
  Sum & 11 & 17 & 28 \\ \bottomrule
   \hline
\end{tabular}
\end{table}

\subsection{Global Effect: Security Issue Density}
\begin{figure}
    \centering
    \includegraphics[width=0.9\linewidth]{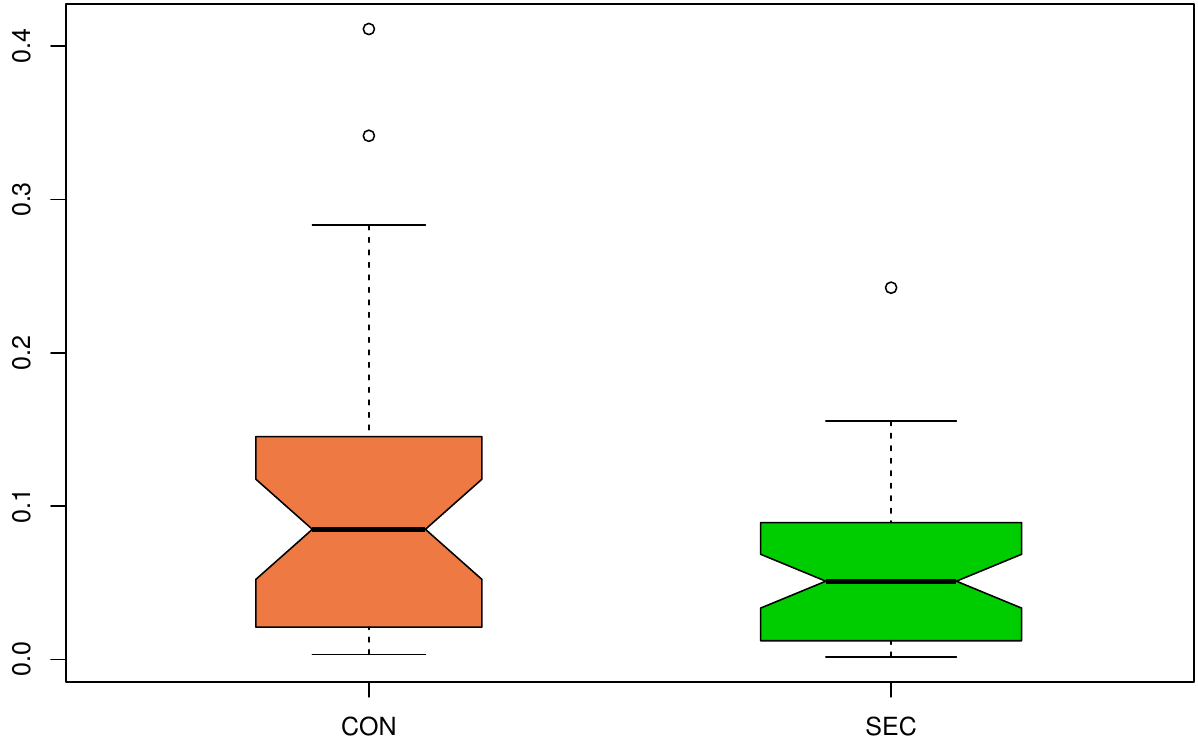}
    \caption{Security issue density number of security issues per lines of code by group. The bold lines indicate the median and the notches the 95\,\% confidence interval of the median. }
    \label{fig:global}
\end{figure}

We start with the main research question, \ie whether incentivizing security improvements leads to lower security issue density compared to equivalent incentives for general code quality. Figure~\ref{fig:global} shows the boxplots of the security issue density of the teams in the control group (CON) and the security-incentivized group (SEC), indicating a clear superiority of the SEC group.

The median of the CON group is 0.08 (CI: 0.05--0.12) and of the SEC group is 0.05 (CI 0.03--0.07). To test for significance, we have to consider that the values are ratios and the distributions are skewed. We therefore applied a Beta-regression model (Table \ref{tbl:beta-regression}), which yielded a significant result ($\beta=-0.396, p=0.0342$). Thus, we could identify a significant group effect pointing in the desired direction, indicating that our security-incentivization mechanism increases code security. 

\begin{table*}
\caption{Security Issue Density: Beta regression coefficients (mean model with logit link) for difference between CON and SEC groups. Notes: The reference category was group=CON, i.e., the control group without incentivation; '.' indicates $p<0.1$, '*' = $p<0.05$, '**' = $p<0.01$, '***' = $p<0.001$.}\label{tbl:beta-regression}
	\centering
	\begin{tabular}{lrrrrc}
		\toprule
		                  & Estimate & Std. Error & $z$ value & $\Pr(>|z|)$  \\ \midrule
		(Intercept)       &  $-2.2223$ &     0.1481 &   $-15.002$ & $<$2e-16 &*** \\
		(grp)SEC &  $-0.3963$ &     0.1871 &   $ -2.118$ & 0.0342 &* \\\bottomrule
	\end{tabular}
\end{table*}

To deepen the understanding of this promising result, we investigated the following aspects that might have affected the teams' performance in the SEC and CON groups: productivity in terms of lines of code (LOC), effort in terms of number of security issues, and efficiency in terms of number of security issues per LOC, \ie security issue density.

\subsection{Productivity: Lines of Code}\label{sec:loc}

\begin{figure}[h!]
	\centering
	\includegraphics[width=0.95\linewidth]{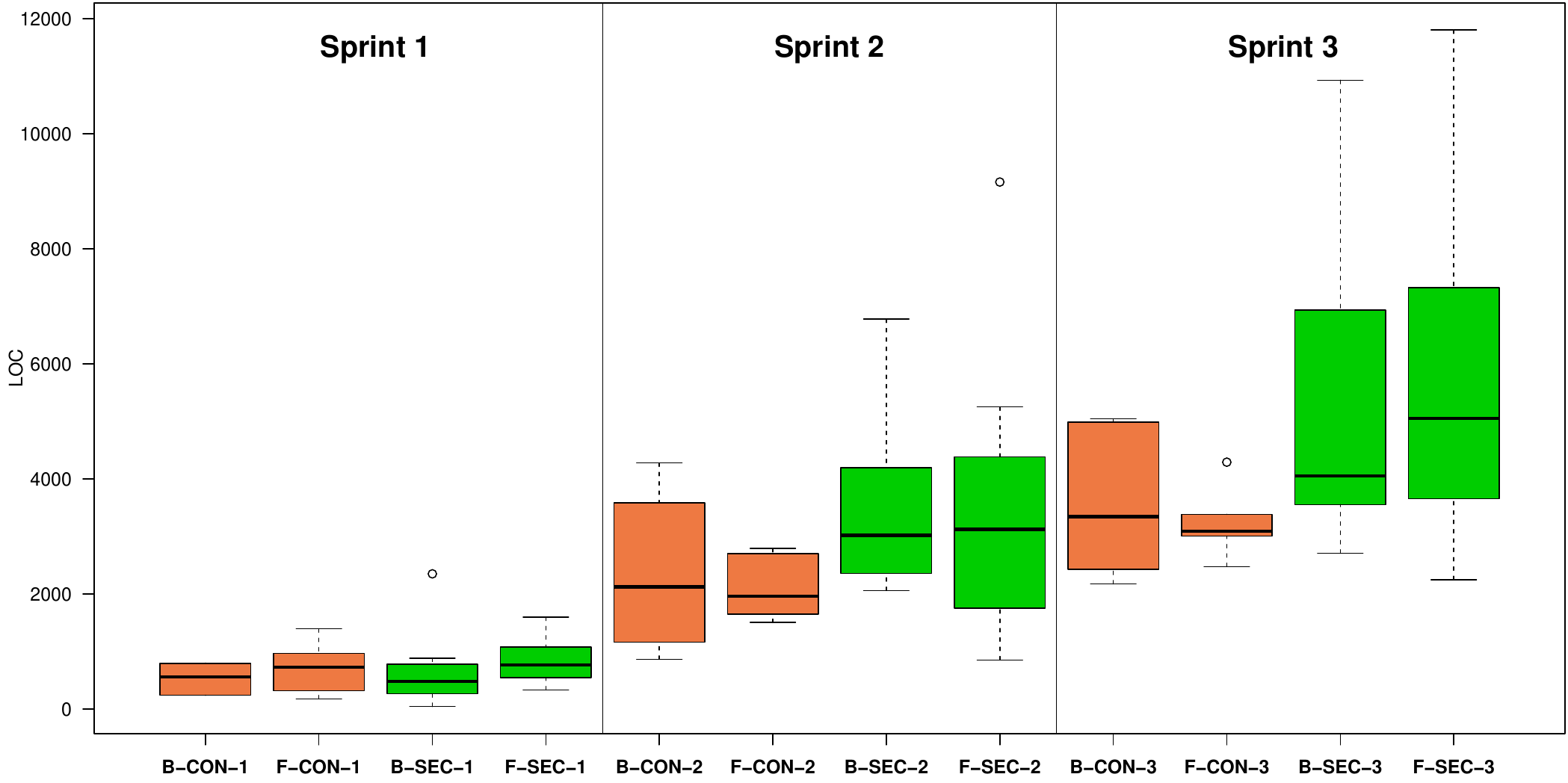} 
	\caption{LOC by group, what, and sprint. Notes: 
           B = back-end, F = front-end; 
           CON = control group,
           SEC = security-incentivized group.}
	\label{fig:loc}
\end{figure}

An indicator of the groups' productivity is the mere number of lines of code written from one sprint to the next. Figure~\ref{fig:loc} shows the LOC by group, layer, and sprint. For an inferential assessment, we chose Poisson regression, as we deal with a count variable. All effects were significant (see Table~\ref{tab:loc}): The Security groups had more code than the control group, the frontend codes were slightly longer than the backend ones, and the LOC increased remarkably at both sprint 2 (four times the first sprint) and 3 (six times the initial value). From the interactions, we learn that the security incentivized groups produced more increase at sprints 2 and 3 than the control groups, indicating a first positive effect of the incentivization.

\begin{table*}
\centering
\caption{Coefficients of the Poisson-regression model of lines of code by group, layer, and sprint.}
\label{tab:loc}
\begin{tabular}{lrrrrl}  \toprule
                                 &Estimate& Std. Error &z value& Pr(>|z|)&    \\\midrule
(Intercept)                      & 6.28071&    0.01766 &355.568&  < 2e-16& ***\\
(what)Frontend                   & 0.29947&    0.02331 & 12.848&  < 2e-16& ***\\
(grp)SEC                         & 0.24469&    0.02225 & 10.995&  < 2e-16& ***\\
(Runde)2                         & 1.48500&    0.01956 & 75.911&  < 2e-16& ***\\
(Runde)3                         & 1.89568&    0.01894 &100.067&  < 2e-16& ***\\
(what)Frontend:(grp)SEC          &-0.09057&    0.02958 & -3.062&   0.0022& ** \\
(what)Frontend:(Runde)2          &-0.41763&    0.02633 &-15.859&  < 2e-16& ***\\
(what)Frontend:(Runde)3          &-0.39707&    0.02533 &-15.673&  < 2e-16& ***\\
(grp)SEC:(Runde)2                & 0.15026&    0.02453 &  6.126& 9.00e-10& ***\\
(grp)SEC:(Runde)3                & 0.16209&    0.02378 &  6.816& 9.36e-12& ***\\
(what)Frontend:(grp)SEC:(Runde)2 & 0.22773&    0.03311 &  6.878& 6.05e-12& ***\\
(what)Frontend:(grp)SEC:(Runde)3 & 0.26465&    0.03192 &  8.291&  < 2e-16& ***\\\bottomrule
\end{tabular}
\end{table*}

\subsection{Effort: Absolute Number of Security Issues}\label{sec:iss}

\begin{figure}
    \centering
    \includegraphics[width=0.95\linewidth]{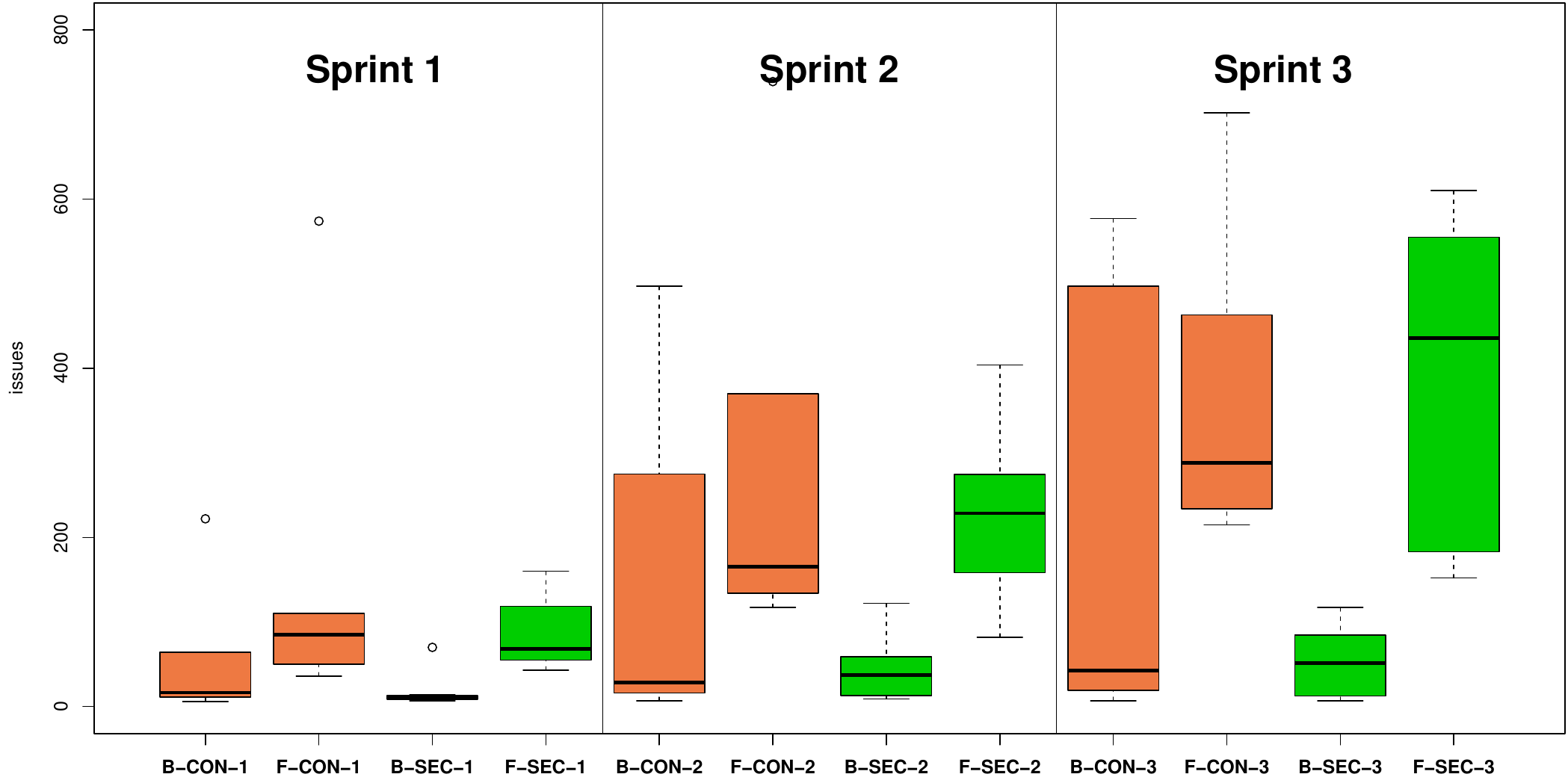}  
    \caption{Security issues by group, what, and sprint. Notes: B = back-end, F = front-end; CON = control group, SEC = security-incentivized group.}
    \label{fig:iss}
\end{figure}

To deepen the quality aspect of the code, we take a look at the number of security issues. 
We first inspect the absolute number of issues by group, part, and sprint (Figure~\ref{fig:iss}).

Figure~\ref{fig:iss} shows several interesting details: First of all, the boxplots indicate highly skewed distributions with medians far below the third quartile. Therefore, the normality assumption of the linear model is implausible. Moreover, as the number of security issues is a count variable, we applied Poisson regression for the significance tests. The results of this test are listed in Table~\ref{tab:iss}.

Second, the number of security issues increased markedly (and significantly) from Sprint 1 to Sprint 2 and remained on a relatively high level after Sprint 3. This may be due to teams first creating a framework for the solution in Sprint 1, which is then filled with more complex functionality prone to inducing issues. Interestingly, the number of security issues remained at a higher level after Sprint 3, particularly in the front-ends developed by teams in both groups. However, these need not necessarily be the same issues as after Sprint 2, but rather new ones due to adding more functionality in Sprint 3 (see Figure~\ref{fig:loc}). 

Third, the front-ends showed (again significantly) more security issues than the back-ends. This might be due to more effort put into the server-side implementation, which (a) is pivotal for the project to work (front-ends are interchangeable) and (b) are server-side issues more detrimental than those on the front-end (a server hack will compromise all users, in contrast to a user's hack).

Fourth, there was a significant interaction of layers (back-end vs. front-end) and group (SEC vs. CON). Inspecting Figure \ref{fig:iss}, we see that the CON groups have roughly the same number of security issues in the front-ends and the back-ends, whereas the SEC groups showed more issues in the front-end than in the back-end. This indicates that the SEC groups were more inclined to deliver clean implementations of their server-side architecture. 

Fifth, there was also a significant interaction of sprint and platform. Inspecting Figure~\ref{fig:iss}, we see that while there were virtually equally few security issues at Sprint 1, the number of issues in the front-ends increased more at Sprints 2 and 3 than those in the back-ends. This supports the assumption that more effort was put into solving security issues in the back-end than in the front-end. 

\begin{table*}
\centering
\caption{Coefficients of the Poisson-regression model of security issues by group, layer, and sprint. Note: Reference categories were  the first for all factors, i.e., CON (grp), Sprint 1, and back-end (what).}
\label{tab:iss}
\begin{tabular}{lrrrl}  \toprule
                                   &Estimate &Std. Error &$z$-value& $\Pr(>|z|)$    \\\midrule
(Intercept)                        & 4.02535 &   0.05455 & 73.786&  $<$ 2e-16 ***\\
(what)front-end                    & 1.02877 &   0.06356 & 16.185&  $<$ 2e-16 ***\\
(grp)SEC                           &-1.15603 &   0.10034 &-11.521&  $<$ 2e-16 ***\\
(sprint)2                          & 0.93048 &   0.06442 & 14.444&  $<$ 2e-16 ***\\
(sprint)3                          & 1.26039 &   0.06181 & 20.392&  $<$ 2e-16 ***\\
(what)front-end:(grp)SEC           & 0.55481 &   0.11220 &  4.945&   7.61e-07 ***\\
(what)front-end:(sprint)2          &-0.34328 &   0.07619 & -4.506&   6.62e-06 ***\\
(what)front-end:(sprint)3          &-0.41461 &   0.07308 & -5.673&   1.40e-08 ***\\
(grp)SEC:(sprint)2                 &-0.02130 &   0.11874 & -0.179&    0.85762    \\
(grp)SEC:(sprint)3                 &-0.16651 &   0.11527 & -1.445&    0.14859    \\
(what)front-end:(grp)SEC:(sprint)2 & 0.40230 &   0.13328 &  3.019&    0.00254 ** \\
(what)front-end:(grp)SEC:(sprint)3 & 1.08843 &   0.12850 &  8.470&  $<$ 2e-16 ***\\\bottomrule
\end{tabular}
\end{table*}

\subsection{Efficiency: Security Issue Density Revisited}\label{sec:ipl}

\begin{figure}
	\centering
	\includegraphics[width=0.95\linewidth]{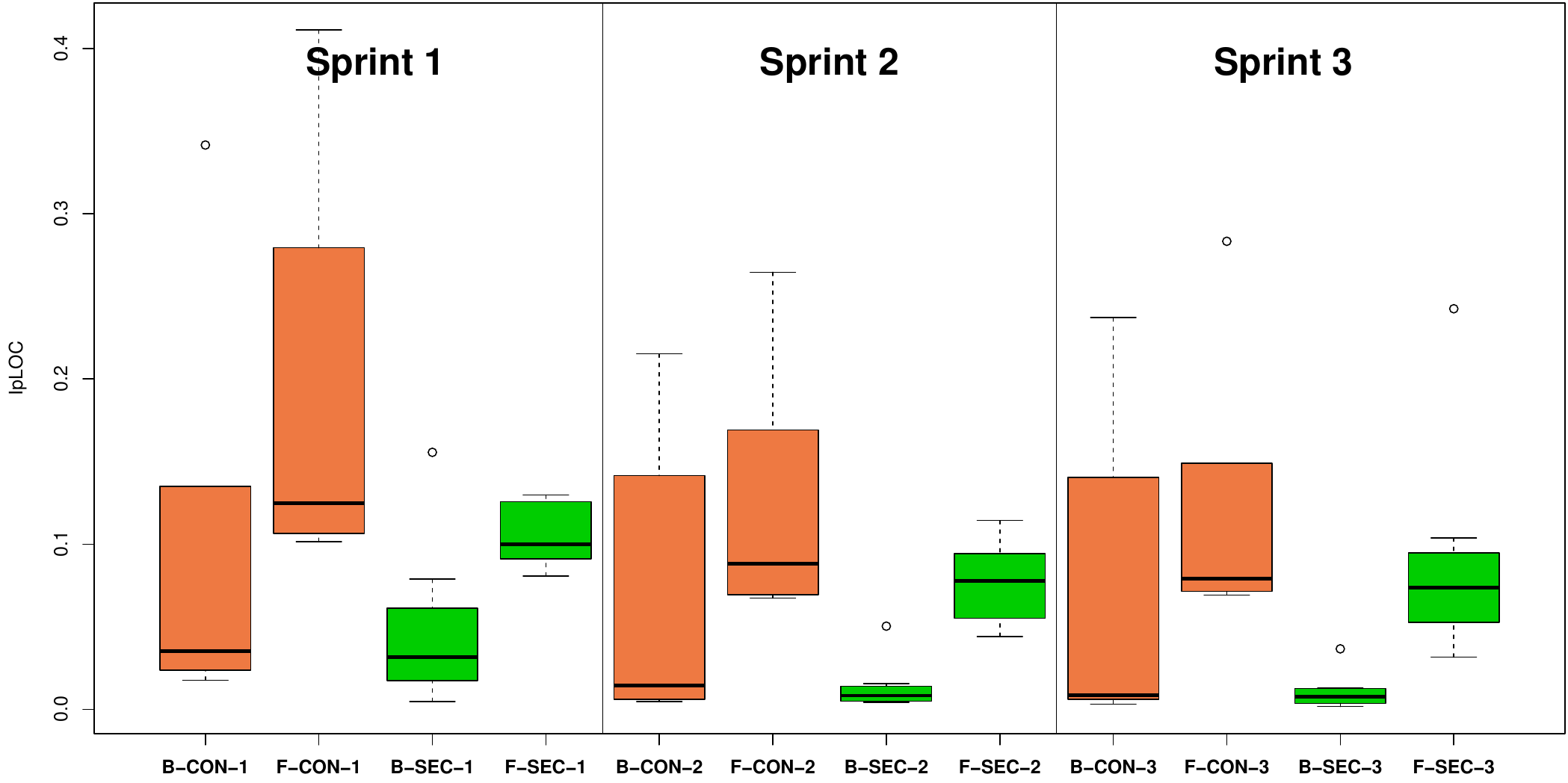}
	\caption{Security issue density by group and sprint. Notes: B = back-end, F = front-end; CON = control group, SEC = security-incentivized group.}
	\label{fig:ipl}
\end{figure}

As the number of security issues is prone to increase with the number of LOC, we explored further the security issue density. Note, this approach does not show the "absolute" efforts the teams put into their projects, which was the subject in Section~\ref{sec:iss}, but rather the effectiveness of their efforts: More LOC require more checking, but was the reduction of security issues proportional to the program length? We applied Beta regression again to address this question. The results are shown in Figure~\ref{fig:ipl}.

We now find a more differentiated picture: The most striking (and significant) difference of defect density is identified between the front-end and the back-end platform (in favor of the back-end). After controlling for this effect, the groups' differences do not achieve significance any longer. However, from Figure~\ref{fig:ipl} we learn that teams in the SEC group had continuously fewer security issues per LOC than teams in the CON group, yet outclassed by the layer effect. Hence, the implications of this analysis step are two-fold: First, the SEC teams put more \textit{successful} effort into the back-end development and they were generally more effective (yet not statistically significant after controlling for all effects). From Figure~\ref{fig:ipl} we learn that the back-ends of the SEC teams had virtually \emph{no} security issues per LOC left after Sprints 2 and 3 whereas several CON teams had up to 0.28 issues per LOC left after Sprints 2 and 3. This result supports our research hypothesis that our incentivization mechanism had indeed the intended effect.

\begin{table*}
\centering
\caption{Coefficients of the Beta-regression model of security issue density by group and sprint. 
         Note: Reference category was the first for all factors.}
\label{tab:ipl}
\begin{tabular}{lrrrl}  \toprule
                                     & Estimate & Std. Error & z value & Pr($>|z|$) \\   \midrule
  (Intercept)                        & -2.51 & 0.29 & -8.63 & 0.00 ***\\ 
  (What)front-end                    &  1.03 & 0.36 &  2.85 & 0.00 **\\ 
  (Grp)SEC                           & -0.43 & 0.40 & -1.08 & 0.28 \\ 
  (Sprint)2                          & -0.58 & 0.44 & -1.30 & 0.19 \\ 
  (Sprint)3                          & -0.70 & 0.45 & -1.56 & 0.12 \\ 
  (What)front-end:(Grp)SEC           & -0.06 & 0.50 & -0.12 & 0.91 \\ 
  (What)front-end:(Sprint)2          &  0.14 & 0.55 &  0.26 & 0.80 \\ 
  (What)front-end:(Sprint)3          &  0.23 & 0.55 &  0.42 & 0.68 \\ 
  (Grp)SEC:(Sprint)2                 & -0.06 & 0.61 & -0.11 & 0.92 \\ 
  (Grp)SEC:(Sprint)3                 & -0.07 & 0.62 & -0.11 & 0.91 \\ 
  (What)front-end:(grp)SEC:(Sprint)2 &  0.19 & 0.76 &  0.25 & 0.81 \\ 
  (What)front-end:(grp)SEC:(Sprint)3 &  0.26 & 0.77 &  0.34 & 0.74 \\    \bottomrule
\end{tabular}
\end{table*}

\subsection{For A Few Points More: The Reward}\label{sec:reward}

\begin{figure}
    \centering
    \includegraphics[width=0.95\linewidth]{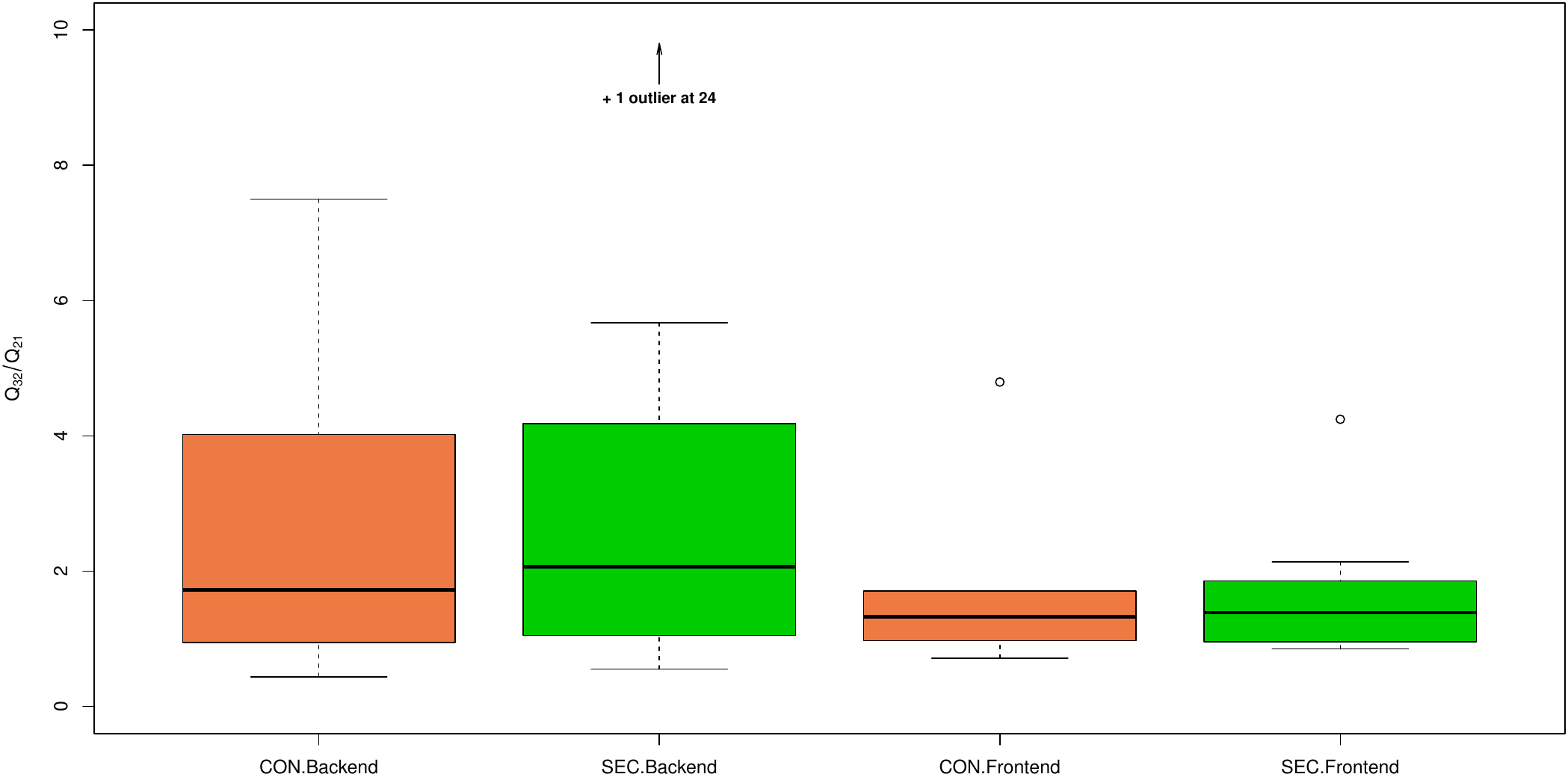}
    \caption{$\Delta Q$ by group and layer (back-end, front-end). }
    \label{fig:DQ}
\end{figure}

To scrutinize the effect of the chosen incentive, we inspect the ratio of security issue density change according to Equation \eqref{eq:issue-density}. Figure~\ref{fig:DQ} shows the boxplots of $\Delta Q$ by group and layer.

The relative change $\Delta Q$ shows no marked difference between the SEC and the CON group, but the ratios are larger for the back-end parts compared to the front-ends. However, none of these differences was significant (Table~\ref{tab:DQreg}). 
\begin{table}
\centering
\caption{$\Delta Q$ by group and layer. Notes: The reference categories were control (CON) and back-end.}
\label{tab:DQreg}
\begin{tabular}{rrrrr} \toprule
                 & Estimate & Std. Error & $t$-value & Pr($>|t|$) \\ \midrule
(Intercept)      &   2.7258 & 1.8514 &  1.47 & 0.1539 \\ 
  SEC            &   2.1808 & 2.4492 &  0.89 & 0.3821 \\ 
  front-end      &  -0.9163 & 2.6183 & -0.35 & 0.7294 \\ 
  SEC:front-end  &  -2.3036 & 3.4637 & -0.67 & 0.5123 \\ \bottomrule
\end{tabular}
\end{table}

Comparing the average of $\Delta Q$ between teams in CON and SEC as shown in Table~\ref{tab:DQx}, we observe about twice as much average change for the back-ends in favor of the SEC group $(4.9:2.7)$. This indicates greater effort that SEC teams put into fixing security issues in their back-ends, supporting our research hypothesis.

\begin{table}
\centering
\caption{Average $\Delta Q$ ratios by group and layer.}
\label{tab:DQx}
\begin{tabular}{lcc} \toprule
    &back-end &front-end\\\midrule
CON &   2.73 &    1.81\\
SEC &   4.91 &    1.69\\\bottomrule
\end{tabular}
\end{table}

\subsection{Answer to the Research Question and Key Findings}  %
Our study demonstrates that incentivizing developers measurably improves security outcomes.
We could identify lower security issue density for the security incentivized group, particularly in the back-end components but also, yet to a lower degree, in the front-end components.  
Generally more effort was put into the back-end components, which showed virtually no security issues in the incentivized group at sprints 2 and 3. 

Key findings of the study hence were:
\begin{itemize}
	\item Security-incentivized teams (SEC) had significantly lower security issue density than the control group (medians 0.05 vs 0.08; $\beta = -0.396, p = 0.0342$ in Beta regression).
	\item Productivity rose similarly across groups (LOC increased from Sprint 1 to 3; no group or layer effects), indicating incentives improved quality rather than code volume.
	\item A strong front-end/back-end disparity emerged; SEC back-ends reached near-zero issue density after sprints 2 and 3, suggesting cleaner server-side implementations.
\end{itemize}

\section{Discussion}

\subsection{Implications for Research}

Our results show a difference between CON and SEC with a significant one for the back-ends. In sum, our findings support our hypothesis that incentivizing developers helps to improve security in software projects. Furthermore, the current implementation of our reward mechanism is applicable to software projects that use Git for version control and a SAST tool to scan code for security issues. Researchers can use our approach and experimental setup to replicate the experiment with other academic and also industrial software projects. Regarding industrial projects, exploring different payoffs for teams to improve security is an interesting direction for future work. Furthermore, our findings indicate potential differences in how security is handled in front-ends and back-ends (cf. Figure \ref{fig:ipl}, speculatively explainable by a usability emphasis in front-end vs. security salience in back-end). Therefore, we view studying this potential difference in more detail as an interesting direction for future research. 

Finally, because our experiment was conducted within a one-semester course, a longitudinal replication that tracks teams through deployment to production would help validate external validity and capture post-course remediations. Such studies could assess whether incentivization maintains its observed benefits at production readiness and refine the timing, structure, and magnitude of incentives to maximize security outcomes.

\subsection{Implications for practitioners}

We operationalized team-level security performance via an automated pipeline that scripts SAST scanners to compute and export issue counts. Given these data, the evaluation of team performance, and the derivation of rewards via \eqref{eq:reward} or under the fairness criteria in Section \ref{sec:reward-assignment} reduces to applying established formulas, for which open-source implementations are available \cite{CoopGame2019} and have been demonstrated in prior work \cite{rass_incentive-based_2023}. Extending these measurements across multiple sprints enables forecasting and early detection of potential security degradation prior to deployment.

In the light of heterogeneity across stack layers, a practical implication for team leads is to tailor individual security objectives and training to the development context (e.g., front-end vs. back-end) while maintaining uniform organizational security standards. Concretely, this can include component-specific targets, rule configurations, and feedback loops that reflect the distinct development emphases without relaxing baseline requirements.

\subsection{Implications for Education and Training}
It was observed that individual engagement varied with students' personal preferences and skill levels. The pronounced differences in security issue density between front-end and back-end development corroborate this heterogeneity. These results suggest that assessments of software quality, including security, should be differentiated according to students' primary focus (front-end versus back-end). We further recommend instituting reciprocal, intra-team quality reviews of security-related work: back-end developers should evaluate the completeness, rigor, and adequacy of security measures implemented by front-end developers, and vice versa. The objective of these cross-checks is not to require either group to identify additional security defects, but to ensure mutual awareness, maintain continuous and constructive oversight, and promote consistent security standards across components without imposing undue burden.

\subsection{Ethical Aspects}\label{sec:pre-study-ethics}\label{sec:experiment-ethics}

Both, the pre-study and the incentive evaluation experiment, were performed with students, therefore, several ethical aspects needed to be considered.

\paragraph{Pre-Study:}
Participation in the pre-study was voluntary, and all data, including code and results, remained confidential. The study did not impact participants' grades and aimed to foster mutual benefits, but without creating a disadvantage for groups who developed their projects in a programming language that the scanners would not cover (here, Java and Kotlin). 

\paragraph{Incentive Evaluation Experiment:}

In this experiment, students were \emph{not} informed that the course was part of a comparative study. This decision was motivated by two considerations. First, awareness of participation in a study can alter behavior, potentially confounding the results. Second, the bonus systems were designed to appear as standard elements of course administration. The control group's bonus system had been established in previous iterations of the course, and the treatment group's system followed an identical structure with only the target metric changed. Moreover, instructors routinely exercise discretion over grading criteria within their sections. The variation between conditions fell within the normal scope of such autonomy.

Neither experimental condition disadvantaged students relative to each other or to historical norms. Both groups could earn identical maximum point totals (200 base points plus up to 20 bonus points), and both received the same security scan reports, differing only in whether those reports were framed as grade-relevant or informational. Bonus points could only improve grades, not reduce them. Student teams who did not engage with the bonus system received the same grades they would have received in its absence. Because the control condition replicated the grading practices of previous years, students in that group experienced no deviation from established expectations.

All data, including source code repositories and assessment results, were handled confidentially. Individual students and teams are not identified in this paper --- results are only reported at the level of teams and layers. 
\section{Threats to Validity}\label{sec:threats}
    
In the following, we discuss several threats that might have affected the validity of the results reported for the incentive evaluation experiment in Section~\ref{sec:results} and how we addressed them.

\subsection{Internal Validity}

Internal validity concerns whether the observed effects can be attributed to the treatment rather than confounding factors.

\paragraph{Instructor effects.}
Four instructors taught across four sections, with imperfect crossing between instructors and conditions: Instructor~2 taught one section in each condition, providing partial control for instructor effects, but Instructor~1 taught only control sections while Instructors~3 and~4 taught only treatment sections. Differences in teaching style, emphasis, or grading strictness could confound the treatment effect. We mitigated this by providing identical written instructions to all sections (Figure~\ref{fig:synopses}) and using automated tools for the primary outcome measurements, but instructor influence on student motivation and behavior cannot be fully controlled.

\paragraph{Cross-contamination.}
Students from the CON and SEC teams may have communicated during the semester, potentially learning about the different bonus criteria. However, the bonus system was not heavily emphasized in either condition; it was presented as one element among many in the course instructions, with the difference amounting to a single word in the course slides (``Sonar scanner'' vs.\ ``security scanner''). Both groups received security scan reports, further reducing the salience of any difference students might have noticed. Furthermore, the same condition applied to all the teams in a course section, limiting information exchange between CON and SEC teams. 
Based on the feedback from all four instructors, no student asked about the slight differences in the synopsis (Figure~\ref{fig:synopses})
 related to security in particular, students could have attributed the difference as subjective for the (different) course instructors, and hence it was (possibly) not further questioned. Hence, we consider cross-contamination unlikely to have substantially affected results, though we cannot rule it out entirely.

\paragraph{Awareness effects.}
Teams in both groups received security scan reports throughout the semester---the teams in the SEC group as part of their bonus system, and the teams in the CON group as informational feedback (``to see some other static analysis tools that might be useful in practice''). This design choice means that both groups were aware that security was being measured, controlling for the possibility that mere awareness of security monitoring (independent of incentivization) drives behavioral change. The key difference between conditions was thus the incentive itself, not the awareness of security concerns.

\paragraph{Gaming of metrics.}
Student teams, who intentionally introduce issues early in the semester only to fix them later could artificially inflate their improvement ratio $\Delta Q$ and earn undeserved bonus points. This threat is inherent to any improvement-based incentive scheme. We did not observe obvious patterns of gaming in the commit histories, but subtle gaming would be difficult to detect. The team-level reward distribution (rather than individual) may reduce gaming incentives, as any individual team member would need cooperation or at least non-objection from other team members.

\paragraph{Assessment timing.}
Assessment dates varied by course section (Monday, Tuesday, or Thursday), but since each section also started on its respective day, all teams had the same amount of development time per sprint. Instructors occasionally granted brief deadline extensions for individual teams at their discretion. We cannot rule out that such extensions were unevenly distributed across conditions, though any effect on the ratio-based outcome metric ($\Delta Q$) is likely minor.

\paragraph{Team size variation.}
Average team size was higher in SEC (6.5 students) than in CON (5.3 students). Larger teams may differ in coordination overhead, individual accountability, or free-rider dynamics, particularly under the team-level bonus structure where all members receive the same reward regardless of individual contribution. We cannot fully disentangle team size effects from the treatment effect, though the overlap in team sizes across conditions (both contained teams of 4--7 members) limits the practical impact of this difference.

\subsection{External Validity}

External validity concerns the generalizability of our findings to other contexts.

\paragraph{Student vs.\ professional developers.}
Our participants were fourth-semester undergraduate students of computer science with limited software development experience and no formal security training. Professional developers may respond differently to security incentives --- they may have stronger baseline security awareness, different motivations (salary vs.\ grades), or face different organizational pressures. Our findings demonstrate that incentivization \emph{can} improve security outcomes in a population with low baseline security knowledge. Whether similar effects hold for experienced professionals remains an open question.

\paragraph{Project domain.}
All projects were networked multiplayer games for Android. This domain was chosen to match the course setting of previous years, but it may not generalize to domains with higher security stakes (e.g., financial or medical software) or different security challenges (e.g., web applications, embedded systems). Games involve authentication and session management but typically do not handle highly sensitive personal data, which may limit the range of security issues encountered.

\paragraph{Incentive type.}
Participants were incentivized with bonus points affecting their course grades --- an immaterial reward with significant personal consequences in an academic context. In professional settings, incentives might be monetary (bonuses, raises), reputational (recognition, promotion), or punitive (consequences for security failures). The effectiveness of incentivization may vary with incentive type and magnitude.

\paragraph{Single institution.}
All data were collected from a single course at one university in Austria. Educational practices, student populations, and cultural attitudes toward security may differ across institutions and countries, limiting generalizability.

\paragraph{Course duration.}
The projects spanned one semester (approximately four months) and did not reach production-ready status. Security concerns often become more acute in later development stages (deployment, maintenance). The observed effects might differ in longer-running projects or in projects that reach production use.

\subsection{Construct Validity}

Construct validity concerns whether our measurements actually capture the constructs of interest.

\paragraph{Security issues vs.\ actual security.}
We measured security using issue counts from static analysis tools (Bearer, Detekt, mobsfscan). These tools detect patterns associated with security vulnerabilities, but SAST findings are not equivalent to actual exploitable vulnerabilities. SAST tools are known to produce false positives (flagged issues that are not real vulnerabilities) and false negatives (real vulnerabilities that are not detected). Our results show that incentivization reduces \emph{tool-reported issues}. Whether this translates to genuinely more secure software requires validation through other means (e.g., penetration testing, dynamic analysis) that we did not perform.

\paragraph{Issue density as a metric.}
We normalized issue counts by lines of code to compute issue density, avoiding penalization of larger codebases. However, this metric assumes that security risk scales linearly with code size, which may not hold. A small amount of security-critical code (e.g., authentication logic) may warrant more attention than a large amount of low-risk code (e.g., UI rendering). The metric also does not account for issue severity, i.e., a single critical vulnerability may matter more than dozens of low-severity warnings.

\paragraph{Multi-tool counting without deduplication.}
We summed issues across three security scanners without removing duplicates, reasoning that multiply-detected issues may be more severe or easier to fix. However, this approach means that tools' overlapping coverage could inflate counts unpredictably, and projects that happen to trigger more overlap would appear to have more issues. A weighted or deduplicated counting scheme might yield different results.

\subsection{Statistical Conclusion Validity}

Statistical conclusion validity concerns the appropriateness of our statistical analyses and the reliability of our conclusions.

\paragraph{Small sample size.}
With only 14 project teams (6 control, 8 treatment), our statistical power to detect effects is limited. Some effects that appear non-significant might achieve significance with larger samples. Conversely, the significant effects we did observe should be interpreted cautiously given the small sample.

\paragraph{Unequal group sizes.}
The treatment group comprised more teams than the control group (8 vs.\ 6 teams, with 52 vs.\ 32 students respectively). This imbalance arose from natural enrollment patterns and section sizes, not from experimental design. Unequal group sizes reduce statistical power and can affect the robustness of some analyses, though the statistical methods we employed (beta regression for issue density and $\Delta Q$, Poisson regression for counts) accommodate unequal groups.

\paragraph{Multiple comparisons.}
We conducted multiple statistical tests across different outcome variables (lines of code, absolute issues, issue density, $\Delta Q$) and breakdowns (by platform, by sprint). This increases the risk of Type~I errors (false positives). We did not apply formal corrections for multiple comparisons (e.g., Bonferroni), so individual significant results should be interpreted with appropriate caution.

\paragraph{Automatic Issue Detection.}
Automatic issue detection is prone to false-positives and -negatives, which we cannot rule out, and manual checks are infeasible. Since the variability is expected to exceed the variability induced by the inclusion/exclusion of auxiliary files by some scanners, we left this out of the analysis. A comparison of issue densities with(out) the auxiliary files included in the lines of code showed that the difference was in the order of second or later digit behind the comma, so it seems safe to ignore it. 
\section{Conclusion}
    We demonstrated that team-level incentives tied to automated security measurements can measurably improve security outcomes: the security-incentivized group (SEC) showed significantly lower issue density than the control group (CON) in the global analysis, without evidence of merely increasing code volume, suggesting quality-oriented effects rather than productivity inflation. After controlling for platform effects, a strong front-end/back-end disparity emerged, with back-ends exhibiting markedly fewer issues; the SEC group's back-end also achieved higher improvement ratios $(\Delta Q)$, indicating that incentives aligned effort toward cleaner server-side implementations. The measurement pipeline, summing findings from Bearer, Detekt, and mobsfscan, proved feasible for scripting and automation, enabling repeatable reporting at team scale. For practitioners, this supports the viability of deploying an automated, largely open tooling chain to quantify and reward team security performance, while tailoring objectives to the distinct emphases of front-end and back-end work. Overall, our results provide promising evidence that incentivization improves measurable security, while motivating follow-up studies in professional settings, with larger samples, extended lifecycles, and more comprehensive validation of actual security impact. \subsection*{Acknowledgment}
We thank Dagmar Auer for invaluable comments to earlier versions of the manuscript.

\begin{acronym}
	\acro{SAST}{Static Application Security Testing}%
	\acro{LOC}{lines of code}%
	\acro{LLM}{large language model}%
	\acro{CC}{Common Criteria}%
	\acro{ROS}{Robot Operating System}%
\end{acronym}

\end{document}